\documentclass[12pt,a4]{article}

\usepackage{color,tikz}
\usepackage[unicode,bookmarks,bookmarksopen,bookmarksopenlevel=2,colorlinks,linkcolor=blue,citecolor=green]{hyperref}

\usepackage{amsmath,eucal,amssymb,amsthm,amsfonts}
\usepackage{mathrsfs,graphicx,texdraw}

\DeclareMathOperator{\diag}{diag}

\def\openone{\leavevmode\hbox{\small1\kern-3.3pt\normalsize1}}
\def\Res{\mathop{\mbox{Res}\,}\limits}

\def\diag{\mbox{diag}\,}

\usepackage[nocompress]{cite}
\def\smath#1{\text{\scalebox{.8}{$#1$}}}
\usepackage{pdflscape}
\newcommand*\diff {\mathop{}\!\mathrm{d}}
\newcommand{\mi}{\mathrm{i}}
\usepackage{tcolorbox}
\usepackage{dsfont}
\usepackage{graphicx}
\usepackage{xcolor}
\usepackage{titlesec}
\usepackage{microtype}

\usepackage{subfig}
\graphicspath{{figures/}} 

\begin{document}

\begin{center}
{\LARGE \bf Nonlocal reductions of the Ablowitz-Ladik \\[6pt]equation}

\bigskip

{\bf Georgi  G. Grahovski \footnote{E-mail: {\tt grah@essex.ac.uk}}, Amal J Mohammed \footnote{E-mail: {\tt ajmoha@essex.ac.uk}} and
Hadi Susanto \footnote{E-mail: {\tt hsusanto@essex.ac.uk}}
}

\medskip

\noindent
{\it  Department of Mathematical Sciences, University of Essex, Wivenhoe Park, Colchester, UK}

\end{center}
\begin{abstract}
\noindent
The purpose of the present paper is to develop the inverse scattering transform for  the  nonlocal semi-discrete nonlinear Schr\"odinger equation (known as Ablowitz-Ladik equation) with ${\cal PT}$-symmetry proposed in \cite{AblMus1}. This includes: the eigenfunctions (Jost solutions) of the associated Lax pair, the scattering data and  the fundamental analytic solutions.
In addition, the paper studies the spectral properties of the associated discrete Lax operator. Based on the formulated (additive) Riemann-Hilbert problem, the 1- and 2-soliton solutions for the nonlocal Ablowitz-Ladik equation are derived.
Finally, the completeness relation for the associated Jost solutions is proved. Based on this, the expansion formula over the complete set of Jost solutions is derived.
This will allow one to interpret the inverse scattering transform as a generalised Fourier transform.

\end{abstract}

\section{Introduction}\label{sec:1}

Completely integrable infinite-dimensional systems are a subject of a constant interest and a large amount  of investigations in different areas of Mathematics and Physics over almost the last five decades \cite{ZMNP,FaTa,GVYa*08} and appear in a wide range of applications - from differential geometry to classical and quantum field theory, fluid mechanics and optics.

A special class of completely integrable infinite-dimensional systems is the class of PDEs integrable by inverse scattering method (ISM) \cite{ZMNP,FaTa}.
The nonlinear Schr\"odinger (NLS) equation
\begin{equation}\label{eq:NLS}
{\rm i}q_t+q_{xx}+2|q^2|q=0, \qquad q=q(x,t),
\end{equation}
appeared at the very early stage of the development of the ISM \cite{ZMNP,FaTa} as one of the classical examples of integrable equations by the ISM and has attracted a significant attention of the scientific community \cite{BMRL2,IP2,ZaSh*74a}. It appears as an universal model for weakly nonlinear dispersive waves,  nonlinear optics and plasma physics \cite{AblPriTru}.

The NLS  model has been generalised in several directions. The first one is to consider multi-component generalisations. The first multi-component/vector generalisation of (\ref{eq:NLS}) was proposed by S. V. Manakov in 1974 (see \cite{ZMNP})
\begin{equation}\label{eq:Man}
i{\bf v}_t+{\bf v}_{xx}+2 ({\bf v}^\dag , {\bf v}) {\bf v}=0,\qquad {\bf v}= {\bf v}(x,t).
\end{equation}
Here ${\bf v}$ is an $n $-component complex-valued vector and $(\cdot ,\cdot) $ is the standard scalar product. It is again integrable by the ISM \cite{AblPriTru,ZMNP,FaTa,GVYa*08}. The two component  VNLS equation (known as the Manakov model) appears in studies of  electromagnetic waves in optical media.
Another direction, motivated by the applications of the differential geometric and Lie algebraic methods to soliton type equations  \cite{vgrn,66,GGMV2011a,GGMV2011b,gc,ForKu*83,2} (for a detailed review see e.g. \cite{GVYa*08})
has lead to the discovery of a close relationship between the multi-component (matrix) NLS equations and the homogeneous
and symmetric spaces~\cite{ForKu*83}.

The first integrable discretisation of the NLS equation (\ref{eq:NLS}) was proposed by M. J. Ablowitz and A. Ladik  and has the form \cite{AL1,AL2,AL3}:
\begin{equation}\label{IDNLS}
  \mi Q_{n,t}=\frac{1}{h^2}(Q_{n+1}-2Q_{n+1}+Q_{n-1})\pm |Q_n|^2(Q_{n+1}+Q_{n-1}),
\end{equation}
It is a differential-difference or semi-discrete equation (discrete in space and continuous in time), and is in fact a $O (h^2)$ finite difference approximation of \eqref{eq:NLS}. The corresponding scattering problem is usually referred to as the Ablowitz-Ladik (AL)
scattering problem \cite{AblPriTru,AblPriTru1,AblBiondPri,Biond,GI*TMF82,GI*DUBNA,GIK*DUBNA80,GIK*JMP84}. The equation \eqref{IDNLS} has also a number of physical applications: it describes  the dynamics of anharmonic lattices \cite{17}, self-trapping on a
dimer \cite{18}, various types of Heisenberg spin chains \cite{19,20} and so on. Later on, various discretisations of the NLS models were studied\cite{BMRL1,BMRL2,AblOhtaTru,TUV2,Veks_Kon,Veks1,Veks2} including perturbation effects \cite{DMR*PRE2003,DMR*PRE2004}.

Recently, in \cite{AblMus,GeSa} was proposed  a nonlocal integrable equation of nonlinear Schr\"odinger type with
${\cal PT}$-symmetry, due to the invariance of
the so-called self-induced potential $V(x,t)=\psi(x,t)\psi^*(-x,-t)$ under the combined action of parity and time
reversal symmetry. In the same paper, the 1-soliton solution for this model is derived and it was shown
that it develops  singularities in finite time. Soon after this, nonlocal ${\cal PT}$-symmetric generalisations are found for
 the Ablowitz-Ladik model in \cite{AblMus1}. All these models are integrable by the Inverse Scattering Method (ISM) \cite{AblMus2}.

The nonlocal reductions of the NLS \eqref{eq:NLS} and the Ablowitz-Ladik equation \eqref{IDNLS} are of particular interest in regards to applications in ${\cal PT}$-symmetric optics, especially in developing of theory of electromagnetic waves in  artificial heterogenic media\cite{Konotop,Barash}. For an up-to-date review, see for example \cite{UFN,Fring2}.

The initial interest in such systems was motivated by quantum mechanics \cite{Bender1, Ali1}. In \cite{Bender1}
it was shown that quantum systems with a non-hermitian Hamiltonian admit states with real eigenvalues, i.e. the hermiticity
of the Hamiltonian is not a necessary condition to have  real spectrum. Using such Hamiltonians one can build  up new quantum mechanics
\cite{Bender1,Bender2, Ali1,Ali2}.  Starting point is the fact that in the case of a non-Hermitian Hamiltonian with real spectrum,
the modulus of the wave function for the eigenstates is time-independent even in the case of complex potentials.

Historically the first pseudo-hermittian hamiltonian with real spectrum is the ${\cal PT}$-symmetric one in \cite{Bender1}. Pseudo-hermiticity here means that
the Hamiltonian ${\cal H}$ commutes with the operators of spatial reflection ${\cal P}$ and time reversal ${\cal T}$: ${\cal P}{\cal T}{\cal H}={\cal H}{\cal P}{\cal T}$.
The action of these operators is defined as follows: ${\cal P}: x\to -x$ and ${\cal T}: t\to -t$. Supposing that the wave function is a scalar, this leads to the following action of the operator of spatial reflection on the space of states: ${\cal P}\psi(x,t)=\psi(-x,t)$ and ${\cal T}\psi(x,t)=\psi^*(x,-t)$. As a result,  the Hamiltonian and the wave function are ${\cal PT}$-symmetric, if ${\cal H}(x,t)={\cal H}^*(-x,-t)$ and $\psi(x,t)=\psi^*(-x,-t)$. Here we used also, that the parity operator ${\cal P}$ is linear and unitary while the time reversal operator ${\cal T}$ is anti-linear and anti-unitary.

The action ${\cal P}$ and ${\cal T}$ operators on the Hamiltonian induces an action on the associated scattering problem (see \eqref{DF1OP} below) and to its potential  \eqref{eq:Lax0}:
\[
  {\cal P} Q_n(t)=Q_{-n}(t), \qquad   {\cal T} Q_n(t)=Q^*_{n}(-t).
\]
This leads to the following reduction (symmetry) condition
\begin{equation}\label{SYMRED}
   Q^-_n(t)=\pm (Q^+)^*_{-n}(t).
\end{equation}
As a result, one gets the nonlocal Ablowinz-Ladik equation with ${\cal PT}$ symmetry, proposed in \cite{AblMus1}:
\begin{equation}\label{SYMRED1}
  \mi Q^+_{n,\tau}=  (Q^+_{n+1}-2Q^+_n+Q^+_{n-1})- \epsilon
  Q^+_n(Q^+)^*_{-n}(Q^+_{n+1}+Q^+_{n-1}), \qquad \epsilon =\pm 1.
\end{equation}
The purpose of this paper is to develop the inverse scattering transform for \eqref{SYMRED1}, to study the spectral properties of the associated Lax operators \eqref{DF1OP} and \eqref{DF2OP} and to derive 1- and 2-soliton solutions.

The structure of the present paper is as follows: in  Section 2 we briefly outline the structure of the semi-discrete Lax representation, the corresponding semi-discrete (differential-difference) zero-curvature equation and the resulting differential-difference equations. In Section 3 we present the direct scattering transform for the nonlocal Ablowitz-Ladik equation. This includes: the Jost solutions, the scattering matrix and the scattering data, and the fundamental analytic solutions.  In Section 4 we formulate a Riemann-Hilbert problem (in additive form) for the fundamental analytic solutions on the continuous spectrum of the discrete Lax operator. Based on this, we derive the 1- and 2-soliton solutions of \eqref{SYMRED1}. finally, in Section 5 we describe the spectral properties of the discrete Lax operator, prove the completeness relation for the Jost solutions and derive expansion formula over the complete set of Jost solutions.

\section{Preliminaries}\label{sec:Prelim}

The starting point here is the semi-discrete analogue of the Lax (or zero-curvature) representation: the initial nonlinear evolutionary equation (\ref{IDNLS}) can be represented as a compatibility condition of two linear systems:
\begin{eqnarray}
   \Psi_{n+1}(z,t)&=&  L_n(z,t) \Psi_n(z,t); \hspace{0.3in} n\in \mathbb{N}, \label{DF1OP}\\
 \Psi_{n,t}(z,t) &=& M_n(z,t)\Psi_n(z,t),\label{DF2OP}
\end{eqnarray}
where
\begin{equation}\label{eq:Lax0}
L_n(z,t)=\left(
             \begin{array}{cc}
               z & Q_n^+(t) \\
               Q_n^-(t) & z^{-1} \\
             \end{array}  \right),
\end{equation}
is an element of the Lie group $SL(2,{\Bbb C})$ and $M_n(z,t)$ is an element of the corresponding Lie algebra $sl (2, {\Bbb C})$. Here we also assume that $Q_n^\pm (t)$ are complex-valued functions, satisfying $\sum_{n=-\infty}^\infty <\infty$. The compatibility condition (i.e. the semi-discrete analogue of the zero-curvature representation) of (\ref{DF1OP}) and (\ref{DF2OP}) takes the form:
\begin{equation}\label{DLax}
  M_{n+1}=  L_{n,t}L_{n}^{-1}+L_nM_nL_{n}^{-1},
\end{equation}
where the operator $M_n$ takes the form $M_n(z,t)=V_n(z,t)+ \Omega(z)$, with
%
\begin{align}\label{Timeeq2}
V_n(z,t)=&\mi \left(
           \begin{array}{cc}
           Q_n^+(Q^+_{1-n})^*  & \hspace{0.3in} z^{-1}Q_{n-1}^+ + zQ_n^+ \\
             &\\
           (z^{-1}(Q^+_{-n})^* - z (Q^+_{1-n})^*) &
              - (Q^+_{-n})^*Q_{n-1}^+ \\
           \end{array}         \right),\qquad
\Omega=&\frac{\mi}{2} (z-z^{-1})^2\sigma_3.
\end{align}
The equation (\ref{IDNLS}) together with appropriate boundary conditions is integrable by the inverse scattering method \cite{AL1,AL2,AL3}. In addition equation \eqref{IDNLS} is one of the members of the integrable hierarchy  associated to the spectral problem \eqref{DF1OP}.

The discrete compatibility condition (\ref{DLax}) results in the following system of differential-difference equations (without using the involution):
\begin{align}\label{COMDIS}
  \mi Q^+_{n,\tau}= & (Q^+_{n+1}-2Q^+_n+Q^+_{n-1})-
  Q^+_nQ^-_n(Q^+_{n+1}+Q^+_{n-1})\nonumber \\
  - \mi Q^-_{n,\tau}= &  (Q^-_{n+1}-2Q^-_n+Q^-_{n-1})-
                     Q^-_n Q^+_n(Q^-_{n+1}+Q^-_{n-1})
\end{align}
If one imposes the standard symmetry condition (involution) $Q_n^-(t)=\epsilon(Q_n^+(t))^*$, then one will get \eqref{IDNLS}. If one sets the nonlocal involution $Q_n^-(t)=\epsilon(Q_{-n}^+(t))^*$, then this will result in \eqref{SYMRED1}.

Finally, we note that the Ablowitz-Ladik Lax operator $L_n(z)$ can be transformed into a spectral (eigenvalue) problem ${\cal L}_n(z)\Psi_n(z)=0$, where
\begin{equation}\label{eq:AL-spec}
{\cal L}_n(z) = \left(\begin{array}{cc}
                        D_+ & 0 \\
                        0 & D_-
                      \end{array}
\right)+ U_n -z{\Bbb I}, \qquad U_n=\left(\begin{array}{cc}
                                            0 & Q_n^+ \\
                                            Q_{n-1}^- & 0
                                          \end{array}
\right).
\end{equation}
Here, $D_{\pm}$ are the shift operators $D_{\pm} \Psi_n(z)=\Psi_{n\pm 1}(z)$ and ${\Bbb I}$ is the ($2\times 2$) identity matrix \cite{GI*TMF82,GI*DUBNA}.

The generic nonlocal involution for the system \eqref{eq:Lax0} can be written in the form
\begin{align}\label{nofinitescatt}
 {\bf C}[L_n(z)]:= B L_{-n}(z^*)^\dagger B^{-1}=L_n(z),
\end{align}
where $B$ is an automorphism of the Lie group $SL(2,{\Bbb C})$.
The particular choice $B=\diag (1,-1)$ leads to \eqref{SYMRED1}.

\section{Direct scattering transform}

\subsection{Jost solutions and scattering data}\label{Green}
The eigenfunctions of $L_n(z)$ and $M_n(z)$ of (\ref{DF1OP})-(\ref{DF2OP}) are defined by their asymptotics (the so-called Jost solutions) for $|n|\to \infty$ (see e.g. \cite{ZMNP,FaTa,GVYa*08}):
\begin{align}\label{DBOC2}
 \psi_n(z)\longrightarrow \left(\begin{array}{cc}
      z^n & \hspace{0.1in} 0 \\
      0 & \hspace{0.1in} z^{-n}
    \end{array}\right),\quad \mbox{as}\quad n\to +\infty,\hspace{0.3in}
    \phi_n(z)\longrightarrow\left(\begin{array}{cc}
     z^n & \hspace{0.1in} 0 \\
      0 & \hspace{0.1in} z^{-n}
    \end{array}\right),\quad \mbox{as}\quad n\to -\infty.
\end{align}
Along with the standard Jost solutions, one can define ``renormalised'' Jost solutions from
the functions $\psi_n(z), \phi_n(z)$, that satisfy
 the scattering problem \eqref{DF1OP}
  \begin{align}\label{assosy}
  \xi_n(z)= \psi_n(z)\textbf{Z}^{-n},\hspace{0.5in}
  \varphi_n (z)= \phi_n(z)\textbf{Z}^{-n}
\end{align}
where $\textbf{Z}=
\left(\begin{array}{cc}
z & 0 \\
0 & z^{-1} \\
\end{array}
\right)
$, and the eigenfunctions $\xi_n,\varphi_n$ are solution of the difference equations respectively

\begin{align}\label{eig12}
     \xi _{n+1}= (\textbf{Z}+ \tilde{\textbf{Q}}_n )\xi_n \textbf{Z}^{-1},\hspace{0.5in}
%
  \varphi_{n+1}= &(\textbf{Z}+ \tilde{\textbf{Q}}_n )\varphi_n \textbf{Z}^{-1},
\end{align}
where $\tilde{\textbf{Q}}_n=
\left(\begin{array}{cc}
0 & Q_n^+ \\
Q_n^- & 0 \\
\end{array}\right)
$, with the canonical  boundary conditions
\begin{equation}\label{bou1}
  \lim_{x\to \infty}\xi_n(z) = {\Bbb I}, \qquad   \lim_{x\to -\infty}\varphi_n(z) = {\Bbb I}.
\end{equation}
The two Jost solutions $\phi_n(z)$ and $\psi_n(z)$ are related by the scattering matrix:
 \begin{equation}\label{scattmatr}
   \phi_n(z)=  \psi_n(z) T ,\hspace{0.3in}
   T=\left(\begin{array}{cc}
     a^{+}(z) & \hspace{0.1in} -b^-(z) \\
      b^{+}(z) & \hspace{0.1in} a^{-}(z)
    \end{array}\right).
 \end{equation}
 The nonlocal involution \eqref{nofinitescatt} imposes symmetry condition as Jost solutions associated with \eqref{eq:Lax0}, and will be
\begin{equation}\label{noDLnM2}
  \textbf{C}(\psi_{-n}^{\dag}((z)^*,t))=B\psi_{-n}^{\dag}((z)^*,t)B^{-1}=
  \phi_n(z,t).
\end{equation}

\subsection{Fundamental Analytic Solutions}\label{scattering a}
Important tools for reducing the ISP to a Riemann--Hilbert problem (RHP) are the fundamental analytic solution (FAS) $\chi^{\pm}
(x,t,\lambda )$. Their construction is based on the  Gauss decomposition of $T(\lambda,t)$, see \cite{ZMNP,58,GG2010}:
\begin{subequations}\label{LIN-FAS-JOS}
\begin{align}
  \chi^+_n(z)= & \psi_n(z)T^-(z)
   =  \phi_n(z)S^+(z),\label{LIN-FAS-JOS1}
\\[2pt]
 \chi^-_n(z)= & \psi_n(z)T^+(z)
   =  \phi_n(z)S^-(z),\label{LIN-FAS-JOS2}
\end{align}
where
\begin{align}\label{INV-SCA}
  %
  T^-(z)=&\left(\begin{array}{cc}
      a^+ (z) \hspace{0.1in}& 0 \\
      b^+ (z)& 1 \\
    \end{array} \right),\hspace{0.2in}
 T^+(z)=\left(\begin{array}{cc}
      1 & -b^-(z) \\
      0 & a^- (z)\\
    \end{array}\right), \\
 S^+(z)=&\left(\begin{array}{cc}
      1 & \beta^-(z) \\
      0 & \alpha^+ (z)\\
    \end{array}\right),\hspace{0.2in}
 \hspace{0.1in} S^-(z)=\left(\begin{array}{cc}
      \alpha^-(z) & 0 \\
      -\beta^+(z) & 1 \\
    \end{array}\right)
\end{align}
\end{subequations}
are the factors in the Gauss decomposition of the associated scattering matrix $T(z)$:
\begin{align}\label{the FAS16}
T(z)= & T^{-}(z) \hat{S}^{+}(z) =T^{+}(z) \hat{S}^{-}(z),
\end{align}
and are expressed in terms of the matrix elements of the scattering matrix $T(z)$ and its inverse
\begin{equation}\label{INV-SCAa}
  \hat{T}(z)= \left(\begin{array}{cc}
                  \alpha^-(z) & \beta^-(z) \\
                  -\beta^+(z) & \alpha^+(z) \\
                \end{array}\right).
\end{equation}
This construction ensures that $\xi_n^\pm(z)$ are
analytic functions of $z$ for $z \in \Omega_\pm$.

On the unit circle $|z|=1$ (i.e., on the continuous spectrum of $L_n(z)$), the two FAS are linearly dependent:
\begin{equation}\label{The Zakharov-Shabat Dressing Method Local1}
  \tilde{\chi}^+_n(z)- \tilde{\chi}^-_n(z)= \tilde{\chi}^-_n(z) \mathbf{G}_n(z) \hspace{0.3in}
  |z|=1,
\end{equation}
 where the sewing function $G_n(z,t)$ can be expressed in terms of $\rho^{\pm}(t,z)$:
 \begin{equation}\label{chijump2}
  \mathbf{G}_n(z,t)= \left(  \begin{array}{cc}
                         \rho^+\rho^- & z^{2n}\rho^- \\
                         z^{-2n}\rho^+ & 0 \\
                       \end{array}\right),
     \hspace{0.4in}\tilde{\chi}^-_n(z)\to \mathds{1} \hspace{0.2in}\text{as}\hspace{0.1in}|z|\to \infty.
\end{equation}
The independent matrix elements of $G_{n}(z,t)$, together with the discrete spectrum of $L_{n}(z)$ form up the  minimal set of scattering data of $L_n$.

The nonlocal involution \eqref{nofinitescatt} imposes symmetry condition on the FAS
and the scattering matrix as follows:
\begin{equation}\label{noFAS}
  \textbf{C}(\chi_{-n}^{\{-,\dag\}}((z)^*,t))=B\chi_{-n}^{\{-,\dag\}}((z)^*,t)B^{-1}=
  \chi^+_n(z,t)
\end{equation}
and
\begin{equation}\label{noSMA}
  \textbf{C}(T^{\dag}((z)^*,t))=BT^{\dag}((z)^*,t)B^{-1}.
\end{equation}
As a result, we obtain
\begin{equation}\label{noRSAM}
  a^{\pm}(z,t)=(a^{\pm}(z^*,t))^*, \qquad b^{\pm}(z,t)=\epsilon(b^{\mp}(z^*,t))^*.
\end{equation}

\subsection{Asymptotic Behavior of FAS}\label{FANf}
Until the end of this section, we will presume that the involution \eqref{nofinitescatt} holds true.

\subsubsection{Asymptotic Behavior of FAS for $|z|=1$}

The discrete scattering problem \eqref{DF1OP} can possess discrete eigenvalues. This can occur when $a^{\pm}(z_j)=0$ for some $z_j$. Here we will assume that this cannot happen on the continuous spectrum. On the zeroes $z_j$ of $a^{\pm}(z)$, the two Jost solutions become proportional:
\begin{equation}\label{Res1}
  \varphi_n^{\pm}(z_j)=\pm b^{\pm}_jz_j^{\mp 2n}\xi_n^{\pm}(z_j).
\end{equation}
Assume that $a^+(z)$ has $S$ simple zeros
$\left\{z_j:|z_j|>1\right\}^S_{j=1}$ and $a^-(z)$ has $S$ simple zeros
 $\left\{z_j:|z_j|<1\right\}^S_{j=1}$, i.e., the number of zeroes inside the unit circle is equal to the number of zeroes outside the unit circle. Then, by equations  \eqref{Res1}:
\begin{align}\label{Res2}
  \text{Res}(\tilde{\varphi}^{\pm}_n,z^{\pm}_j)= & \frac{\varphi^{\pm}_n(z^{\pm}_j)}{\dot{a}^{\pm}(z^{\pm}_j)}
  =\pm \frac{b^{\pm}_j (z^{\pm}_j)^{\mp 2n} \xi_n^{\pm}}{\dot{a}^{\pm}(z^{\pm}_j)} =\pm(z^\pm_j)^{\mp 2n}C^{\pm}_{j}
  \xi^{\pm}_n(z^{\pm}_j),
\end{align}
where we have denoted the norming constants as $C^{\pm}_{j}$.

\subsubsection{Asymptotic Behavior of FAS for $|z|\to \infty$ and $|z|\to 0$}

Then the Laurent expansions of FASs $\chi^{\pm}_n(z)$ have the following form
\begin{subequations}\label{LauraFAS}
  \begin{align}
    \chi^+_n(z)= &
    \left(\begin{array}{cc}
     1+O(z^{-2},\text{even}) &  \hspace{0.2in}   c_n^{-1} z^{-1}Q^+_{n}+O(z^{-3},\text{odd})
     \vspace{0.1in} \\
     z^{-1}Q^-_{n-1}+O(z^{-3},\text{odd}) &  \hspace{0.1in}   c_n^{-1}+O(z^{-2},\text{even})\\
    \end{array}\right)  \hspace{0.2in} \text{as}\ |z| \to \infty,\label{LauraFAS1} \\
    & \nonumber\\
    \chi^-_n(z)= &
    \left(  \begin{array}{cc}
       c_n^{-1}+ O(z^{2},\text{even})&\hspace{0.2in} zQ^+_{n-1}+O(z^{3},\text{odd}) \vspace{0.1in}\\
        c_n^{-1}zQ^-_{n}+ O(z^{3},\text{odd}) &\hspace{0.1in} 1+O(z^{2},\text{even}) \\
      \end{array}  \right) \hspace{0.2in} \text{around}\hspace{0.3in} z=0.\label{LauraFAS2}
  \end{align}
\end{subequations}
From the analytic properties of the eigenfunctions of $L_n(z)$, it follows that $a^+(z)$ has an analytic extension in the region $|z|\to \infty$:
\begin{subequations}\label{analx}
  \begin{align*}
    W(\varphi^+_n,\xi^+_n)= &W
     \left( \begin{array}{cc}
           1+O(z^{-2},\text{even})  & c_n^{-1}z^{-1}Q_n+O(z^{3},\text{odd})\vspace{0.1in}  \\
           z^{-1}R_{n-1}+O(z^{-3},\text{odd})  & c_n^{-1}+O(z^{-2},\text{even})  \\
           \end{array}\right),
  \end{align*}
  then $\chi^+_n(z)$ is analytic when $|z|\to \infty$
  \begin{equation}\label{analx+}
    a^+(z)=1-O(z^{-2},\text{even})\hspace{0.2in} \text{as}\ |z|\to \infty.
  \end{equation}
 Similar arguments apply in order to find  the Laurent series expansions for $a^{-}_n(z)$:
 \begin{align*}
    W(\xi^-_n,\varphi^-_n)= &W
     \left( \begin{array}{cc}
           c_n^{-1}+O(z^{2},\text{even}) & zQ_{n-1}+O(z^{3},\text{odd}) \vspace{0.1in}\\
           c_n^{-1}zR_{n}+O(z^{3},\text{odd}) & 1+O(z^{2},\text{even}) \\
           \end{array}\right),
  \end{align*}
so, $\chi^-_n(z)$ is analytic around $|z|= 0$ and
  \begin{equation}\label{analx-}
    a^-(z)=1-O(z^{2},\text{even})\hspace{0.2in} \text{as}\ |z|\to 0.
  \end{equation}
\end{subequations}
The scattering coefficients can be written as infinite explicit sums of the eigenfunctions:
\begin{align}\label{forsca}
  a^+(z)= & 1+ \sum_{k=-\infty}^{+\infty} z^{-1} Q_k^+ \varphi^{(2),+}_k,\hspace{0.3in}
  b^+(z)=   \sum_{k=-\infty}^{+\infty}z^{2k-1}Q^-_k\varphi^{(1),+}_k,\\
  a^-(z)= & 1 + \sum_{k=-\infty}^{+\infty} z Q_k^- \varphi^{(1),-}_k\hspace{0.2in},\hspace{0.3in}
   b^-(z)=  - \sum_{k=-\infty}^{+\infty}z^{-2k+1}Q^+_k\varphi^{(2),-}_k,
\end{align}

\section[The Riemann-Hilbert] {The Riemann-Hilbert Problem and Soliton Solutions}\label{RHP}

It is well known that the inverse scattering transform for the Lax operator $L_n(z)$ can be reduced to a Riemann-Hilbert boundary value problem on the complex plane. The contour, where the values of the analytic functions are specified is the continuous spectrum of $L_n(z)$ \eqref{DF1OP} - in the case of the Ablowitz-Ladik equation it is the unit circle $|z|=1$. If the Lax operator $L_n(z)$ has discrete eigenvalues, then the resulting Riemann-Hilbert problem (RHP) is of singular type. Here we will restrict ourselves to the so-called balanced RHPs -- we will consider only problems having equal number of singularities inside and outside the boundary contour, i.e. we will assume that the number of zeroes of $a^+(z)$ is equal to the number of zeroes of $a^-(z)$. The reflectionless case $b^\pm (z)=0$ will correspond to soliton solutions of \eqref{SYMRED1} with $\epsilon=-1$.

\begin{itemize}
\item \text{\textit{Symmetries and symmetry reductions}}\\

Since the expansions of $a^{\pm}(z)$ we presented in Section \ref{scattering a}  contain only even power of $z^{-1}$ and $z$, it follows that  if $z^{\pm}_j$ are zeroes of $a^{\pm}(z)$ then $-z^{\pm}_j$  are also zeros of $a^{\pm}(z)$. This implies that
\begin{equation}\label{SYMre1}
  C^{\pm}_j(-z^{\pm}_j)=  \frac{b^{\pm}(-z^+_j)}{\dot{a}^{\pm}(-z^{\pm}_j)}=C^{\pm}(z^+_j),
\end{equation}
where $b^-_j=-b^+_j$, and
\begin{equation}\label{SYMre3}
  \rho^+(-z)=-\rho^+(z), \hspace{0.2in} \rho^-(-z)=-\rho^-(z).
\end{equation}
The nonlocal involution \eqref{nofinitescatt} will act on the reflection coefficient and the normalisation constant producing the following constraint:
\begin{equation}\label{noRETR}
  C^-_j(z_j^-)=\frac{\epsilon(b^+(z_j^{+,*}))^*}{(\dot{a}^-(z_j^{-,*})^*},
  \qquad
  \rho^-(z)=\frac{\epsilon(b^+(z^*))^*}{(a^-(z^*))^*}.
\end{equation}
\item \text{\textit{ Case of poles}}\\
If the FAS $\tilde{\varphi}^{\pm}_n(z)$ have poles, the method of solution of the RHP  requires an extra step involving a contour integration. The starting point is the relations between the eigenfunctions,
 %
 \begin{align}
  \tilde{\varphi}^{\pm}_n=\frac{\varphi^{\pm}_n}{a^{\pm}}(z)= &\xi^{\mp}_{n}(z)\pm
  z^{\mp 2n}\rho^{\pm}(z) \xi^{\pm}_{n}(z) . \label{NOpolelincom1}
\end{align}
We will apply the contour integration method to the following integral representations:
\begin{subequations}
 \begin{align}
  \mathcal{J}_{1,n}(z) = & \frac{1}{2\pi \mi}
  \left(\oint_{\gamma^+}\frac{\diff \omega \varphi^+_{n}(\omega)}{(\omega-z)a^+(\omega)}-
  \oint_{\gamma^-} \frac{\diff \omega \xi^-_{n}(\omega)}{(\omega-z)}\right),\label{NONCON1}
  \\
  \mathcal{J}_{2,n}(z) = & \frac{1}{2\pi \mi}
  \left(\oint_{\gamma^+} \frac{\diff \omega \xi^+_{n}(\omega)}{(\omega-z)} -
  \oint_{\gamma^-}\frac{\diff \omega \varphi^-_{n}(\omega)}{(\omega-z)a^-(\omega)}\right)
  , \label{NONCON2}
\end{align}
\end{subequations}

where the contours of integration $\gamma_\pm$ are depicted in Figure 1. Here we will present the detailed evaluation of one of the integrals ($\mathcal{J}_{2,n}$). The other one can be evaluated in a similar manner.

\begin{figure}[htp]
\centering
\includegraphics[scale= 0.64]{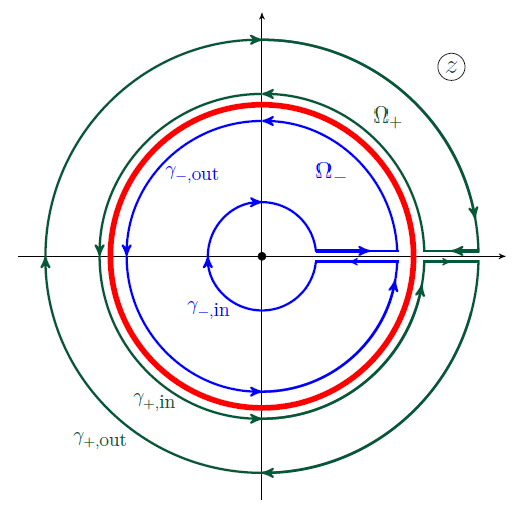}\label{fig:IST}
\caption{The continuous spectrum $\Omega$  of $L_{n}(z)$ (in red) and the integration contours}
\end{figure}

Recall that $(\mi) \ \frac{1}{a^+(z)}$ has simple pole at $z=z^+_j$ ;
$(\mi \mi) \ \frac{1}{a^-(z)}$ has simple pole at $z=z^-_j$ and
$(\mi \mi \mi) \ \xi^-_{n}, \xi^+_{n}$ have no poles, and therefore the integrand of
the first integral in $\mathcal{J}_{2,n}(z)$ has only a pole at $z=\omega$
and ($\mi$v) outside the contour is a negatively oriented,
while inside the contour is a positively oriented.
Thus when $z\in \Omega_+$, we found:
\begin{equation}\label{NONCON3}
  \mathcal{J}_{2,n}(z) = \xi^+_{n}(z)-
  \sum_{j=1}^{S} \left[ \frac{\varphi^-_{n}(z^-_j)}{(z-z^-_j)\dot{a}^-_j}+
  \frac{\varphi^-_{n}(-z^-_j)}{(z+z^-_j)\dot{a}^-_j}\right],
\end{equation}
 and the integral along the unit circle $\Omega={z\in {\Bbb C}\,:\, |z|=1}$ is  equal to
 \begin{equation}\label{NONCON5}
   \frac{1}{2\pi \mi}
  \left(\oint_{\gamma^+} \frac{\diff \omega \xi^+_{n}(\omega)}{(\omega-z)} -
  \oint_{\gamma^-}\frac{\diff \omega \varphi^-_{n}(\omega)}{(\omega-z)a^-(\omega)}\right)
  =\frac{1}{2\pi \mi}
 \oint_{\textcolor[rgb]{0.00,0.00,0.55}{\Omega}} \frac{\diff \omega }{(\omega-z)}\ \omega^{2n}
 \rho^-(z)\xi^-_{n}(z).
 \end{equation}
 \begin{subequations}\label{NONCON78}
If $a^{\pm}(z)$ have zeros at $z^{\pm}_j$ respectively, then one can write
 \begin{equation}\label{zerolinear}
   \varphi_n^{\pm}=\pm (z_j^{\pm})^{\mp 2n} b^{\pm}_j \xi^{\pm}_n,
 \end{equation}
 and this will lead to the following integral representation for $\xi^+_{n}(z)$
 \begin{align}\label{NONCON7}
   \xi^+_{n}(z)=\binom{0}{1}+\frac{1}{2\pi \mi}
 \oint_{\textcolor[rgb]{0.00,0.00,0.55}{\Omega}} \frac{\diff \omega }{(\omega-z)}\
 \omega^{2n}\rho^-(\omega)\xi^-_{n}(\omega)-
 \sum_{j=1}^{S} C^-_j (z^-_j)^{2n}\left[ \frac{\xi^-_{n}(z^-_j)}{(z-z^-_j)}+
  \frac{\xi^-_{n}(-z^-_j)}{(z+z^-_j)}\right].&
 \end{align}
In a similar way, one can find an integral representation for $\xi^-_{n}(z)$
with $z\in \Omega_-$ from evaluating the integral $\mathcal{J}_{1,n}$,
\begin{align}\label{NONCON8}
   \xi^-_{n}(z)=\binom{1}{0}+\frac{1}{2\pi \mi}
 \oint_{\textcolor[rgb]{0.00,0.00,0.55}{\Omega}} \frac{\diff \omega }{(\omega-z)}\
 \omega^{-2n}\rho^+(\omega)\xi^+_{n}(\omega)+
 \sum_{j=1}^{S} C^+_j (z^+_j)^{-2n}\left[ \frac{\xi^+_{n}(z^+_j)}{(z-z^+_j)}+
  \frac{\xi^+_{n}(-z^+_j)}{(z+z^+_j)}\right].&
 \end{align}
\end{subequations}
 The integrands in the  integral representations \eqref{NONCON7} and \eqref{NONCON8} are expressed in terms of $\rho^\pm (z)$ on the continuous spectrum only, while the sums that give the contributions from the discrete spectrum are expressed in terms of the normalisation constants $C_j^\pm(z_j^\pm)$. Thus the system of singular integral
 equations \eqref{NONCON78} admits unique solution, so  the minimal set of scattering data
$
\mathcal{F}_1=\{\rho^+(z),\rho^-(z),z\in |z|=1,z^{\pm}_j,j=1,\dots S\}
$
 contains all information needed to recover uniquely  the Jost solutions $ \xi^{\pm}_{n}(z)$.
 \end{itemize}
 \subsection{1-Soliton solution}

In the case where the Lax operator $L_n(z)$ comprises proper eigenvalues but $\rho^+(z)=\rho^-(z)=0$ on
the continuous spectrum of $L_n(z)$, then  \eqref{NONCON78} reduces to linear system of algebraic
equations:
\begin{subequations}\label{NONCON91}
  \begin{align}\label{NONCON9}
   \xi^+_{n}(z)=& \binom{0}{1}-
 \sum_{j=1}^{S} C^-_j (z^-_j)^{2n}\left[ \frac{\xi^-_{n}(z^-_j)}{(z-z^-_j)}+
  \frac{\xi^-_{n}(-z^-_j)}{(z+z^-_j)}\right],\\
  \xi^-_{n}(z)=& \binom{1}{0}+
 \sum_{j=1}^{S} C^+_j (z^+_j)^{-2n}\left[ \frac{\xi^+_{n}(z^+_j)}{(z-z^+_j)}+
  \frac{\xi^+_{n}(-z^+_j)}{(z+z^+_j)}\right].\label{NONCON10}
 \end{align}
\end{subequations}
 Putting $z=\pm z_j^\pm $ in \eqref{NONCON9} and \eqref{NONCON10}, respectively gives
\begin{subequations}\label{DNLSCON}
  \begin{align}
   \xi^+_{n}(\pm z^+_j)=& \binom{0}{1}\mp
 \sum_{k=1}^{S} C^-_k (z^-_k)^{2n}\left[ \frac{\xi^-_{n}(z^-_k)}{(z^+_j\mp z^-_k)}+
  \frac{\xi^-_{n}(-z^-_k)}{(z^+_j \pm z^-_k)}\right],\label{DNLSCON1}\\
%
 %
  %
  \xi^-_{n}(\pm z^-_j)=& \binom{1}{0}\pm
 \sum_{k=1}^{S} C^+_k (z^+_k)^{-2n}\left[ \frac{\xi^+_{n}(z^+_k)}{(z^-_j\mp z^+_k)}+
  \frac{\xi^+_{n}(-z^+_k)}{(z^-_j\pm z^+_k)}\right],\label{DNLSCON3}
  %
 \end{align}
\end{subequations}
so, the above relations show that
\begin{subequations}\label{RESY}
\begin{align}
  \xi^{+,1}_n(-z^+_j) \hspace{0.09in}& =  -\xi^{+,1}_n(z^+_j),\hspace{0.3in} \text{iff}
  \hspace{0.2in} \xi^{-,1}_n(-z^-_j)  =  \hspace{0.08in}\xi^{-,1}_n(z^-_j), \label{RESY1}\\
 \xi^{+,2}_n(-z^+_j)\hspace{0.09in} & =\hspace{0.09in} \xi^{+,2}_n(z^+_j),\hspace{0.35in} \text{iff}
  \hspace{0.2in} \xi^{-,2}_n(-z^-_j)  =\hspace{0.03in}-\xi^{-,2}_n(z^-_j).\label{RESY2}
\end{align}
\end{subequations}
Then it follows that the solution of \eqref{NONCON9} and \eqref{NONCON10} reads
\begin{align}
  \xi^{-,1}_n(z^-_1)= & \left[1-4C^+_1C^-_1\frac{(z_1^+)^{-2(n-1)}(z^-_1)^{2n}}{((z^+_1)^2-(z^-_1)^2)^2}\right]
  ^{-1},\label{SLSYSTEM1}\\
  \xi^{+,2}_n(z_1^+)= & \left[1+4C^+_1C^-_1\frac{(z_1^-)^{2(n-1)}(z^+_1)^{-2n}}{((z^+_1)^2-(z^-_1)^2)^2}\right]
  ^{-1}.\label{SLSYSTEM2}
\end{align}
Moreover, the potentials are given by
\begin{eqnarray}
  Q^+_{n-1} &=& -2C_1^+(z_1^+)^{-2n-2}\xi_n^{+,(2)} (z_1^+),\label{POTEQ1}\\
  Q^-_n &=&  2C_1^-(z_1^-)^{2n-2}\xi_n^{-,(1)} (z_1^-).\label{POTEQ2}
\end{eqnarray}
Then substituting equation \eqref{SLSYSTEM1} and \eqref{SLSYSTEM2} in \eqref{POTEQ1} and \eqref{POTEQ2} will give  the general form of 1-soliton  solution
\begin{align}
  Q^+_{1n}= & -\frac{ 2 C_1^- (z^-_1)^{2n}}{1+4C^+_1C^-_1((z^+_1)^2-(z^-_1)^2)^{(-2)}(z_1^+)^{-2n}(z^-_1)^{2(n+1)}}
  \label{One Soliton1},\\
  Q^-_{1n}= & \frac{ 2 C_1^+ (z^+_1)^{-2(n+1)}}{1+4C^+_1C^-_1((z^+_1)^2-(z^-_1)^2)^{(-2)}(z_1^+)^{-2n}(z^-_1)^{2(n+1)}}
  \label{One Soliton2}.
\end{align}
%
%
Taking into account the time evolution of the  normalisation constants:
\begin{equation*}
  C^+_1(z,\tau) = C^+_1(0)e^{2\mi \omega^+_1 \tau} ,\hspace{0.2in}
 C^-_1(z,\tau) = C^-_1(0)e^{-2\mi \omega^-_1 \tau},
\end{equation*}
where $\omega^{\pm}=\frac{\mi}{2}(z^{\pm}_1-(z^{\pm}_1)^{-1})$ and the canonical symmetry condition $Q_n^- = -Q_{-n}^*$, one can reduce \eqref{One Soliton1} and \eqref{One Soliton2} into the standard form of the 1-soliton solution of  nonlocal AL equation:
\begin{equation}\label{DNNLSEs}
Q^+_{1n}=\frac{((z_1^{\smath{+}})^2-(z_1^{\smath{-}})^2)
e^{\mi \alpha_1^-}e^{-2\mi \omega^-_1 \tau}(z_1^{\smath{-}})^{2n-1}}
  {1+(z_1^-)(z_1^+)^{-1}e^{\mi (\alpha^+_1 + \alpha^-_1)}e^{2\mi (\omega^{\smath{+}}_1-\omega^{\smath{-}}_1)\tau}(z_1^{\smath{+}})^{-2n}(z_1^{\smath{-}})^{2n}}.
\end{equation}
This reproduces the result obtained by M. Ablowitz and Z. Musslimani.

\subsection{2-Soliton solution}
In a similar way  as before (for the one pole) we will obtain the two-soliton solutions from a singular RHP with quartet of discrete eigenvalues/singularities $z_{\{1,2\}}^+$ and $z_{\{1,2\}}^-$.
%
%
%
The starting point here are  the following linear integral equations for $\xi^{\pm}_n(z)$:
\begin{subequations}\label{Lin-Sys}
 \begin{align}
 \xi^+_n(z)= & \binom{0}{1}+\frac{1}{2\pi \mi}
 \oint_{\textcolor[rgb]{0.00,0.00,0.55}{\Omega}} \left(\frac{\diff \omega }{(\omega-z)}\
 (\omega)^{2n}\rho^-(\omega)\xi^-_{n}(\omega)\right)-\nonumber\\
  %
  & \nonumber\\
  & \left[
   C_1^- (z^-_1)^{2n} \left( \frac{ \xi^-_{n}(z^-_1)}{(z-z^-_1)}
  + \frac{
   \xi^-_{n}(-z^-_1)}{(z+z^-_1)}\right) \right]+
  \left[C^-_2 (z^-_2)^{2n}  \left( \frac{ \xi^-_{n}(z^-_2)}{(z-z^-_2)}
  + \frac{
  \xi^-_{n}(-z^-_2)}{(z+z^-_2)}\right) \right],\label{Lin-Sys1}
  \end{align}
  \begin{align}
  \xi^-_n(z)= & \binom{1}{0}-\frac{1}{2\pi \mi}
 \oint_{\textcolor[rgb]{0.00,0.00,0.55}{\Omega}} \left(\frac{\diff \omega}{(\omega-z)}\
 (\omega)^{-2n}\rho^+(\omega)\xi^+_{n}(\omega)\right)+\nonumber\\
   %
  & \nonumber\\
  & \left[
   C_1^+ (z^+_1)^{-2n} \left( \frac{ \xi^+_{n}(z^+_1)}{(z-z^+_1)}
  + \frac{
   \xi^+_{n}(-z^+_1)}{(z+z^+_1)}\right) \right]+
  \left[C^+_2 (z^+_2)^{-2n}  \left( \frac{ \xi^+_{n}(z^+_2)}{(z-z^+_2)}
  + \frac{
  \xi^+_{n}(-z^+_2)}{(z+z^+_2)}\right) \right].\label{Lin-Sys2}
\end{align}
\end{subequations}
The two-soliton solution again correspond to zero reflection coefficients (i.e. $\rho^+(z)=\rho^-(z)=0$ on $|z|=1$). In this case the system \eqref{Lin-Sys} reduces to a linear system of algebraic equations for $\xi^+_n(z)$
\begin{subequations}\label{Lin-Alag}
 \begin{align}
  \xi^+_n(z)= & \binom{0}{1}-
   \left[
   C_1^- (z^-_1)^{2n} \left( \frac{ \xi^-_{n}(z^-_1)}{(z-z^-_1)}
  +\frac{
   \xi^-_{n}(-z^-_1)}{(z+z^-_1)}\right) \right]-\nonumber\\
  & \hspace{2in}\left[C^-_2 (z^-_2)^{2n}  \left( \frac{ \xi^-_{n}(z^-_2)}{(z-z^-_2)}
  + \frac{
  \xi^-_{n}(-z^-_2)}{(z+z^-_2)}\right) \right],\label{Lin-Alag1}
 \end{align}
and a similar system for $\xi^-_n(z)$:
 \begin{align}
  \xi^-_n(z)= & \binom{1}{0}+
  \left[
   C_1^+ (z^+_1)^{-2n} \left( \frac{ \xi^+_{n}(z^+_1)}{(z-z^+_1)}
  + \frac{
   \xi^+_{n}(-z^+_1)}{(z+z^+_1)}\right) \right]+\nonumber\\
  & \hspace{2in}
  \left[C^+_2 (z^+_2)^{-2n}  \left( \frac{ \xi^+_{n}(z^+_2)}{(z-z^+_2)}
  + \frac{
  \xi^+_{n}(-z^+_2)}{(z+z^+_2)}\right) \right].\label{Lin-Alag2}
\end{align}
\end{subequations}
Here $\xi^+_n(\smath{\pm} z^+_{j})$ are the FAS $\xi^+_n(z)$ evaluated at the eigenvalue
$\smath{\pm} z^+_{j}$ (similarly, $\xi^-_n(\smath{\pm} z^-_j)$ are the FAS $\xi^-_n(z)$
evaluated at the eigenvalue $\smath{\pm} z^-_{j}$).
We can find the expressions for these vectors by evaluating \eqref{Lin-Alag1} at the points
$\pm z_{\{1,2\}}^+$ and  \eqref{Lin-Alag2} at the points $\pm z_{\{1,2\}}^-$.
This results in a linear algebraic system composed of \eqref{Lin-Alag1} and \eqref{Lin-Alag2}
\begin{subequations}\label{Lin-Alag-Sy1}
 \begin{align}
  \xi^+_n(\pm z_1^+)= & \binom{0}{1}-
   \left[
   C_1^- (z^-_1)^{2n} \left( \frac{ \xi^-_{n}(z^-_1)}{(\pm z_1^+-z^-_1)}
  +\frac{
   \xi^-_{n}(-z^-_1)}{(\pm z_1^+ + z^-_1)}\right) \right]-\nonumber\\
  & \hspace{2in}\left[C^-_2 (z^-_2)^{2n}  \left( \frac{ \xi^-_{n}(z^-_2)}{(\pm z_1^+-z^-_2)}
  + \frac{
  \xi^-_{n}(-z^-_2)}{(\pm z_1^+ + z^-_2)}\right) \right],\label{Lin-Alag3}\\
  \xi^+_n(\pm z_2^+)= & \binom{0}{1}-
   \left[
   C_1^- (z^-_1)^{2n} \left( \frac{ \xi^-_{n}(z^-_1)}{(\pm z_2^+-z^-_1)}
  +\frac{
   \xi^-_{n}(-z^-_1)}{(\pm z_2^+ + z^-_1)}\right) \right]-\nonumber\\
  & \hspace{2in}\left[C^-_2 (z^-_2)^{2n}  \left( \frac{ \xi^-_{n}(z^-_2)}{(\pm z_2^+-z^-_2)}
  + \frac{
  \xi^-_{n}(-z^-_2)}{(\pm z_2^+ + z^-_2)}\right) \right],\label{Lin-Alag4}
  \end{align}
\end{subequations}
\begin{subequations}\label{Lin-Alag-Sy2}
 \begin{align}
  \xi^-_n(\pm z_1^-)= & \binom{1}{0}+
  \left[
   C_1^+ (z^+_1)^{-2n} \left( \frac{ \xi^+_{n}(z^+_1)}{(\pm z_1^- - z^+_1)}
  + \frac{
   \xi^+_{n}(-z^+_1)}{(\pm z_1^- + z^+_1)}\right) \right]+\nonumber\\
  & \hspace{2in}
  \left[C^+_2 (z^+_2)^{-2n}  \left( \frac{ \xi^+_{n}(z^+_2)}{(\pm z_1^- -z^+_2)}
  + \frac{
  \xi^+_{n}(-z^+_2)}{(\pm z_1^- +z^+_2)}\right) \right],\label{Lin-Alag5}\\
  &\nonumber\\
  \xi^-_n(\pm z_2^-)= & \binom{1}{0}+
  \left[
   C_1^+ (z^+_1)^{-2n} \left( \frac{ \xi^+_{n}(z^+_1)}{(\pm z_2^- - z^+_1)}
  + \frac{
   \xi^+_{n}(-z^+_1)}{(\pm z_2^- + z^+_1)}\right) \right]+\nonumber\\
  & \hspace{2in}
  \left[C^+_2 (z^+_2)^{-2n}  \left( \frac{ \xi^+_{n}(z^+_2)}{(\pm z_2^- -z^+_2)}
  + \frac{
  \xi^+_{n}(-z^+_2)}{(\pm z_2^- +z^+_2)}\right) \right].\label{Lin-Alag6}
\end{align}
\end{subequations}
From \eqref{Lin-Alag-Sy1} and  \eqref{Lin-Alag-Sy2}, it follows that
\begin{subequations}\label{Rel-Con-Sy}
\begin{align}
  \xi^{+,1}_n(-z^+_j) \hspace{0.09in}& =  -\xi^{+,1}_n(z^+_j) \hspace{0.3in} \text{iff}
  \hspace{0.2in} \xi^{-,1}_n(-z^-_j)  =  \hspace{0.08in}\xi^{-,1}_n(z^-_j), \label{Rel-Con-Sy1}\\
 \xi^{+,2}_n(-z^+_j)\hspace{0.09in} & =\hspace{0.09in} \xi^{+,2}_n(z^+_j) \hspace{0.35in} \text{iff}
  \hspace{0.2in} \xi^{-,2}_n(-z^-_j)  =\hspace{0.03in}-\xi^{-,2}_n(z^-_j).\label{Rel-Con-Sy2}
\end{align}
\end{subequations}
We can recover $Q_n^-$ from the power series expansion of the RHS of $\xi_n^{-,2}(z)$ in
\eqref{Lin-Alag2},
%
and take the residue at $z \to z^+_1$ or at $z \to z^+_2$,  from which we then can obtain
%
%
  \begin{subequations}\label{Pote1-2pol}
  \begin{equation}
   Q_n^-= -\frac{1}{2}C^+_1 (z^+_1)^{-2n-2}\xi^{+,2}_{n}(z^+_1)+
  \frac{2C^+_2 (z^+_2)^{-2n} }{((z^+_1)^2-(z^+_2)^2)}\xi^{+,2}_{n}(z^+_2),\label{Pote1-2pol1}
  \end{equation}
%
or
  \begin{equation}
  Q_n^-=\frac{2C^+_1 (z^+_1)^{-2n} }{((z^+_2)^2-(z^+_1)^2)}\xi^{+,2}_{n}(z^+_1)
  -\frac{1}{2}C^+_2 (z^+_2)^{-2n-2}\xi^{+,2}_{n}(z^+_2).\label{Pote1-2pol2}
  \end{equation}
  \end{subequations}
%
%
%
However it is difficult to find the potential from $\varphi_n^-(z)$.
To fix the problem we will multiply the Laurent expansion function $\chi_n^-(z)$ by
$\left(   \begin{array}{cc}
     1 & 0 \\
     0 & c_n \\
   \end{array} \right)
$
and compare it with RHS of \eqref{Lin-Alag1}
\begin{equation}\label{App-Pot}
  (\tilde{\xi}^-_n,\tilde{\varphi}^-_n)\simeq \tilde{\chi}^-_n(z)=
  \left(\begin{array}{cc}
    1 & 0 \\
    0 & c_n \\
  \end{array} \right)
  \left(  \begin{array}{cc}
       c_n^{-1}&\hspace{0.2in} zQ^+_{n-1} \vspace{0.1in}\\
        c_n^{-1}zQ^-_{n}  &\hspace{0.1in} 1 \\
  \end{array}  \right)\hspace{0.2in} =\hspace{0.2in}
  \left(\begin{array}{cc}
      c^{-1}_n & \hspace{0.2in}zQ^+_{n-1} \\
      zQ_n^- & c_n \\
  \end{array} \right),
\end{equation}
where $\tilde{\chi}_n^-(z)$ has the same power series expansions as $\chi^-_n(z)$. Then we can find the potential $Q^+_{n-1}$ from $\tilde{\varphi}^{-,1}_n$ in \eqref{App-Pot},
 since from \eqref{NOpolelincom1} (when $b^-(z)=0$)
\begin{align}\label{Oth-eig}
  %
  \tilde{\varphi}^-_n(z) = & \binom{0}{1}-
   \left[
   C_1^- (z^-_1)^{2n} \left( \frac{ \xi^-_{n}(z^-_1)}{(z-z^-_1)}
  +\frac{
   \xi^-_{n}(-z^-_1)}{(z+z^-_1)}\right) \right]-\nonumber\\
  & \hspace{2in}\left[C^-_2 (z^-_2)^{2n}  \left( \frac{ \xi^-_{n}(z^-_2)}{(z-z^-_2)}
  + \frac{
  \xi^-_{n}(-z^-_2)}{(z+z^-_2)}\right) \right].
\end{align}
We have a quartet of eigenvalues $\pm z^+_j
, \pm z^-_j
$
with $|z^+_j
|>1$ and $|z^-_j
|<1$ respectively, thus we can solve the linear algebra system \eqref{Lin-Alag3}-\eqref{Lin-Alag6}
for $\xi^-_n(z^-_1)$,$\xi^-_n(z^-_2)$,$\xi^+_n(z^+_1)$ and $\xi^+_n(z^+_2)$.
In particular, in order to find $Q_n^+$ we will  need $\xi^-_n(z^-_1)$ and $\xi^-_n(z^-_2)$ :
\begin{subequations}\label{Soli2solu}
  \begin{align}
  \xi^{+,1}_n(z_1^+)= &  -2\ z_1^+\
  \left[\frac{C_1^- (z^-_1)^{2n}}{((z_1^+)^2-(z^-_1)^2)}\xi^{-,1}_{n}(z^-_1)+
   \frac{C_2^- (z^-_2)^{2n}}{((z_1^+)^2-(z^-_2)^2)}\xi^{-,1}_{n}(z^-_2) \right], \\
  &\nonumber\\
  \xi^{+,1}_n(z_2^+)= &  -2\ z_2^+\
  \left[\frac{C_1^- (z^-_1)^{2n}}{((z_2^+)^2-(z^-_1)^2)}\xi^{-,1}_{n}(z^-_1)+
 \frac{C_2^- (z^-_2)^{2n}}{((z_2^+)^2-(z^-_2)^2)}\xi^{-,1}_{n}(z^-_2) \right],\\
  &\nonumber\\
  \xi^{-,1}_n(z_1^-)= & 1+\frac{2C_1^+(z_1^+)^{-2(n-1)}}{((z_1^-)^2-(z_1^+)^2)}\xi^{+,1}_n(z_1^+)+
  \frac{2C_2^+(z_2^+)^{-2(n-1)}}{((z_1^-)^2-(z_2^+)^2)}\xi^{+,1}_n(z_2^+),
        \label{SYST-EQE1}\\
&\nonumber\\
\xi^{-,1}_n(z_2^-)= & 1+\frac{2C_1^+(z_1^+)^{-2(n-1)}}{((z_2^-)^2-(z_1^+)^2)}\xi^{+,1}_n(z_1^+)+
\frac{2C_2^+(z_2^+)^{-2(n-1)}}{((z_2^-)^2-(z_2^+)^2)}\xi^{+,1}_n(z_2^+),\label{SYST-EQE2}
\end{align}
\end{subequations}
Now, by comparing the power series expansion of the RHS of \eqref{Oth-eig}
to the expansion \eqref{App-Pot}, we obtain the potential when $z \to z_1^-$ or $z \to z_2^-$
%
%
\begin{equation}\label{Pote2-2pol1}
 Q_{n-1}^+ = \frac{1}{2} C_1^- (z^-_1)^{2n-2}\xi^{-,1}_{n}(z^-_1) -
  \frac{2\ C_2^- (z^-_2)^{2n}}{((z^-_1)^2-(z^-_2)^2)}\xi^{-,1}_{n}(z^-_2).
\end{equation}
or
\begin{equation}\label{Pote2-2pol2}
 Q_{n-1}^+ = \frac{-2\ C_1^- (z^-_1)^{2n}}{((z^-_2)^2-(z^-_1)^2)}\xi^{-,1}_{n}(z^-_1)+
  \frac{1}{2} C_2^- (z^-_2)^{2n-2}\xi^{-,1}_{n}(z^-_2).
\end{equation}
%
%
Then substituting each \eqref{SYST-EQE1} and \eqref{SYST-EQE2}, in \eqref{Pote2-2pol1} or \eqref{Pote2-2pol2} and using the involution \eqref{SYMRED}, with negative sign, then equation \eqref{Pote2-2pol1} is the solution for the nonlocal discrete NLS equation.
\section{Spectral properties of $L_n(z)$ and completeness of the Jost solutions}

The crucial fact that determines the spectral properties of the operator
$L_n (z)$ is the choice of the class of functions where from we shall choose
the potential $Q(x) $. Here we assume that $Q_n(t)$ is a differentiable function for all $t\in {\Bbb R}$ and exists for all $n\in {\Bbb Z}$. In additional, we assume that it tends to zero as $n\to \pm \infty$.

The FAS $\chi ^\pm_n (z)$ of $L_n(z)$ allow one to construct
the resolvent $R_{n,m}(z)$ of the operator $L_n (z)$ and then to investigate its spectral
properties. From the general theory of linear operators  we
know that the point in the complex $z$-plane is
a regular point if $R_{n,m}(z)$ is a bounded integral operator.
In each connected subset of regular points $R(z) $ is
analytic in $z$.
The points  which are not regular constitute the spectrum of
$L_n(z) $: the continuous spectrum of $L_n(z) $ consists of all points
$z$ for which $R_{n,m}(z) $ is an unbounded integral operator while the discrete spectrum of $L_n(z)$ consists of all points
$z$ for which $R_{n,m}(z)$ develops pole singularities.

The kernel of the resolvent $R_{n,m}(z)$ can be expressed through
the FAS of $\chi^\pm_n(z)$ as follows:
\begin{subequations}\label{DSpuR}
\begin{align}\label{DSpuR1}
        R^+_{\{n,m\}}(z)= & \chi^+_{n+1}(z)
    \left( \begin{array}{cc}
        \theta(m-n) & 0 \\
        0 &  \theta(n-m)\\
    \end{array} \right)
    \hat{\chi}^+_m(z),\\
        R^-_{\{n,m\}}(z)= & \chi^-_{n+1}(z)
    \left( \begin{array}{cc}
        \theta(n-m) & 0 \\
        0 &  \theta(m-n) \\
    \end{array} \right)
    \hat{\chi}^-_m(z). \label{DSpuR2}
\end{align}
\end{subequations}
 Note that by construction, $R_{\{n,m\}}^{\pm}(z)$ is analytic in $\gamma^{\pm}$ respectively.

We will derive the completeness relation for the Jost solutions of $L_{n}$, by construction the partition of unity of the group of fundamental solutions of $L_n(z)$. For this purpose, we will use again the contour integration method for suitably chosen contours which do not cross the continuous spectrum of $L_n(z)$. We need to evaluate the integral
 \begin{align}\label{ComplR1}
   \mathcal{J}_{R,\{n,m\}}(z) =  \frac{1}{2\pi \mi}
  \left(\oint_{\gamma^+}\diff z \ R^+_{n}(z)-\oint_{\gamma^-}\diff z\  R^-_{n}(z)\right),
 \end{align}
 along the contours $\gamma_\pm$, as shown on Figure 1. According to Cauchy's residue theorem, one has

 \begin{align}\label{ComplR}
   \mathcal{J}_{R,\{n,m\}}(z)= \sum_{j=1}^{N}\left(\Res_{z=\smath{\pm} z^{\smath{+}}_j}R^+_n(z)+
    \Res_{z=\smath{\pm} z^{\smath{-}}_j}R^-_n(z) \right).
 \end{align}
Here we denoted by $z_j^\pm$ the discrete eigenvalues of $L_n(z)$ that are outside/inside the unit circle $|z|=1$, respectively. We will again assume that their numbers are equal and that all of them are isolated singularities for the resolvent.

Each of the integrals in \eqref{ComplR1} can be written as a sum of integral over the continuous spectrum $\Omega$ and an integral involving asymptotics for $|z|\to \infty$ and $|z|\to 0$:
\begin{align}\label{ComplR2a}
\frac{1}{2\pi \mi}
\oint_{\gamma^\pm}\diff z \ R^\pm_{n}(z)=  \int_{|z|=1}\diff z \ R^\pm_{n}(z)-\oint_{\gamma^\pm_{\rm as}}\diff z \ R^\pm_{n}(z)
 \end{align}
First, in order to find the residues in \eqref{ComplR} one needs the Laurent series expansions of the FAS and scattering data around the points of the discrete spectrum $z_j^\pm$. Using \eqref{LIN-FAS-JOS}, one can  write the following expansions:
\begin{subequations}\label{RDissp}
  \begin{align}
  a^{\smath{\pm}}(z)= & (z-(\smath{\pm} z^{\smath{\pm}}_j))\dot{a}^{\smath{\pm}}_j
  +\frac{1}{2} (z-(\smath{\pm} z^{\smath{\pm}}_j))^2 \ddot{a}^{\smath{\pm}}_j + \dots ,
  \label{RDissp1}\\[0.5pt]
  \alpha^{\smath{\pm}}(z)= & (z-(\smath{\pm} z^{\smath{\pm}}_j))\dot{\alpha}^{\smath{\pm}}_j
  +\frac{1}{2} (z-(\smath{\pm} z^{\smath{\pm}}_j))^2 \ddot{\alpha}^{\smath{\pm}}_j + \dots ,
  \label{RDissp1a}\\[0.5pt]
  \chi^+_n(z^+_j)&= \psi^+_{n,j}(z)(b^+_j,1)= \phi^+_{n,j}(z)(1,1/b^+_{j}),   \label{RDissp2}\\[1pt]
  \hat{\chi}^+_n(z^+_j)&= \binom{1}{-\beta^+_j}
  \frac{\tilde{\Psi}^+_{n,j}(z)}{(z-\smath{\pm}z_j^{\smath{+}})
  \dot{\alpha}_j^{\smath{+}}}=
  \binom{1/\beta^+_j}{-1}
  \frac{\tilde{\Phi}^+_{n,j}(z)}{(z-\smath{\pm}z_j^{\smath{+}})\dot{\alpha}_j^{\smath{+}}},
  \label{RDissp3}\\[0.5pt]
  \chi^-_n(z^-_j)&= \psi^-_{n,j}(z)(1,-b^-_j)= \phi^-_{n,j}(z)(-1/b^-_{j},1) ,  \label{RDissp4}\\[1pt]
  \hat{\chi}^-_n(z^-_j)&= -\binom{\beta^-_j}{1}
  \frac{ \tilde{\Psi}^-_{n,j}(z)}{(z-\smath{\pm}z_j^{\smath{-}})\dot{\alpha}_j^{\smath{-}}}=
  \binom{1}{1/\beta^-_j}
  \frac{\tilde{\Phi}^-_{n,j}(z)}{(z-\smath{\pm}z_j^{\smath{-}})\dot{\alpha}_j^{\smath{-}}},
  \label{RDissp5}
  \end{align}
\end{subequations}
where $\tilde{\Psi}_n(z)$ and $\tilde{\Phi}_n(z)$ are related with $\psi_n(z)$ and $\phi_n(z)$ respectively.
As a result, one can find the residues  of $R^{\pm}_n(z)$ at $z=\smath{\pm}z^{\pm}_j$:
\begin{equation}\label{ResKer}
 \Res_{z=\smath{\pm}z^{\pm}_j}R^{\pm}_{n,m}(z) = \mp\frac{\phi^{\pm}_{n+1,j}(z) \tilde{\Psi}^{\pm}_{m,j}(z)}
 {\dot{\alpha}_j^{\smath{{\pm}}}(z)}.
\end{equation}
Now, we can calculate the jump of $R_n(z)$ on the unite circle $(|z|=1)$. Using \eqref{DSpuR1}
one can obtain the result
 \begin{equation}\label{DRJum3}
\int_{|z|=1}\diff z \,( R^+_{n,m}(z)-R^-_{n,m}(z))=
   \int_{|z|=1}\diff z \,\left(\frac{ \phi^+_{n+1}(z) \tilde{\Psi}^+_{m}(z)}{\alpha^{\smath{+}}(z)}
   +\frac{\phi^-_{n+1}(z) \tilde{\Psi}^-_{m}(z)}{\alpha^{\smath{-}}(z)}
   \right).
 \end{equation}
Finally, in order to calculate the contribution of the integrals over the asymptotic circles (i.e when $|z|\to \infty$ and $|z|\to 0$, respectively). For this purpose we need the asymptotics of FAS $\chi^\pm_{n,as}(z)$ for $z \to \infty$ and $z\to 0$:

\begin{equation*}
  \chi^{+}_{\{{\rm as},n\}}(z)=  \left( \begin{array}{cc}
                 z^n & 0 \\
                 0 & z^{-n} \\
               \end{array} \right)+\mathcal{O}(1/z),\hspace{0.2in} z\to \infty;
  \hspace{0.2in}
  \chi^{-}_{\{{\rm as},n\}}(z)=  \left( \begin{array}{cc}
                 z^n & 0 \\
                 0 & z^{-n} \\
               \end{array} \right)+\mathcal{O}(z),\hspace{0.2in} z\to 0.
\end{equation*}
A direct contour integration shows (if we take a limit $|z|\to \infty$ in the integral over $\gamma_{\rm as}^-$ and a  limit $|z|\to 0$ in the integral over $\gamma_{\rm as}^+$, respectively) that
\begin{equation}\label{DeltCont3}
  \frac{1}{2\pi \mi}\left(\oint_{\gamma^+_{\rm as}}\diff z R^{\smath{+}}_{\{{\rm as},n,m\}}
  +\oint_{\gamma^-_{\rm as}}\diff z R^{\smath{-}}_{\{{\rm as},n,m\}}\right)= \delta(n-m){\Bbb I}.
\end{equation}
As a result, combining \eqref{ComplR1} and \eqref{ComplR} and taking into account \eqref{ResKer}, \eqref{DRJum3} and \eqref{DeltCont3}, one gets the following  completeness relation:
 \begin{align}\label{Gene-Four-COM2}
 \delta(n-m)\mathds{1} =& \frac{1}{2\pi \mi}\int_{|z|=1}\diff z
 \left(\frac{ \phi^+_{n+1}(z) \tilde{\Psi}^+_{m}(z)}{\alpha^{\smath{+}}(z)}
 +\frac{\phi^-_{n+1}(z) \tilde{\Psi}^-_{m}(z)}{\alpha^{\smath{-}}(z)}
 \right) \nonumber\\
 &+  \sum_{j=1}^{N}\left(\frac{\phi^+_{\{n+1,j\}}(z) \tilde{\Psi}^+_{m,j}(z)}
 {\dot{\alpha}_j^{\smath{+}}(z)}-\frac{\phi^-_{\{n+1,j\}}(z) \tilde{\Psi}^-_{m,j}(z)}
 {\dot{\alpha}_j^{\smath{-}}(z)}-\right).
\end{align}
Therefore, the Jost solutions $\phi_{n}^\pm(z)$ form a complete set of functions over the space of fundamental solutions of $L_n(z)$.

Based on the completeness relation \eqref{Gene-Four-COM2}, one can expand every function $Y(x)$ from the space of solutions of $L_n(z)$ over the complete set Jost solutions by the following expansion formulae:
%
\begin{align}\label{Gene-Four-COM3a}
  Y(x)= & \frac{1}{2\pi \mi}\int_{|z|=1}\diff z
 \left(\phi^+_{n+1}(z)y^+_{m} (z)
 +\phi^-_{n+1}(z) y^-_{m}(z)
 \right) \nonumber\\
 &+  \sum_{j=1}^{N}\left(\phi^+_{\{n+1,j\}}(z) y^+_{\{m,j\}}(z)
 -\phi^-_{\{n+1,j\}}(z) y^-_{\{m,j\}}(z)\right),
 \end{align}
%
where
\begin{subequations}\label{Gene-Four-COM3b}
  \begin{align}
  y^{\pm}_m(z)= & \frac{1}{\alpha^{\pm}(z)} \int_{|z|=1} \diff z\  \tilde{\Psi}^{\pm}_m(z) Y_m(z),\label{Gene-Four-COM3ba}\\
  y^{\pm}_{\{m,j\}}= & \frac{1}{\dot{\alpha}^{\pm}_j}
  \int_{|z|=1} \diff z\  \tilde{\Psi}^{\pm}_{\{m,j\}}(z)Y_m(z). \label{Gene-Four-COM3bb}
\end{align}
\end{subequations}
 %
%
\section{Conclusions}\label{sec:concl}

We have studied here a nonlocal version \cite{AblMus1} of the semi-discrete NLS equation in the Ablowitz-Ladik form. This equation appears to be ${\cal PT}$ symmetric.  We formulated the direct scattering problem for the nonlocal Ablowitz-Ladik equation. This incudes: the  construction of  the Jost solutions and the minimal set of scattering data, the construction of the fundamental analytic solutions (FAS). Then, based on the formulation of the inverse scattering transform for \eqref{SYMRED} in the form of additive Riemann-Hilbert boundary value problem, the 1- and 2-soliton solutions are derived.

It was shown in \cite{AblMus1} that the 1-soliton solution develops a singularity in finite time. This was due to the disbalance of the associated Riemann-Hilbert problem - the numbers of zeroes of the FAS inside the boundary contour is not equal to the number of zeroes inside the contour: the nonlocal involution requires that if $z_j$ is a discrete eigenvalue then $z_j^*$ must me also an eigenvalue, i.e.  both $z_j$ and $z_j^*$ must be either inside or outside the unit circle. Depending on the positions of the discrete eigenvalues $z_j^\pm$ in the spectral plane, there are two regimes for the 2-soliton solution: if one of the discrete eigenvalues is inside the unit circle and the other is outside, then the nonlocal involution will preserve their number inside and outside the contour balanced and as a result, the corresponding 2-soliton solutions will be regular for all $t$. Otherwise, the 2-soliton solution will develop again a singularity in finite time.

Finally, we outlined briefly the spectral properties of the Lax operator $L_n(z)$. We have derived the completeness relations for the Jost solutions and obtained expansions over the complete set of Jost solutions for a generic function  from the space
of solutions of $L_n (z)$.

The results of this paper can be extended in several directions:
\begin{itemize}

\item To construct gauge covariant formulation of the inverse scattering method for nonlocal Ablowitz-Ladik equation \eqref{One Soliton1}, including the generating (recursion) operator \cite{G*86} and it spectral decomposition \cite{GG2010}, the description of the class of the differential-difference equations solvable by the spectral problem \eqref{DF1OP} (i.e. the corresponding integrable hierarchy), the description of the infinite set of
integrals of motion and the hierarchy of Hamiltonian structures.

\item To study the gauge equivalent systems \cite{66,GGMV2011a,GGMV2011b}.

\item To study the inverse scattering method for the equivalent spectral eigenvalue problem \eqref{eq:AL-spec}. In this case the corresponding Riemann-Hilbert problem is with a canonical normalisation.

\item To study the associated Darboux transformations and their generalizations for both local and nonlocal Ablowitz-Ladik equations. This will provide an algebraic method for constructing  and classification of possible soliton solutions, including also  rational solutions \cite{DokLeb}.

\item To extend the results of this paper for the case of non-vanishing boundary conditions (a non-trivial
background) \cite{AblMus3,Li1,Li2,PriVit}.  In the local case, such solutions are of interest in nonlinear optics: they arise
in the theory of ultrashort femto-second nonlinear pulses in optical fibers. The nonlocal reduction of the Ablowitz-Ladik equation can be of particular interest in the  theory
of electromagnetic waves in artificial heterogenic media \cite{UFN}. The considerations required in this
case are more complicated and will be discussed elsewhere.

  \item To study multi-component generalisations \cite{ForKu*83,gc,58,vgrn,Metin} for both local and nonlocal semi-discrete NLS equation. This includes the block Ablowitz-Ladik system \cite{GI*DUBNA} and generalisations to homogeneous and symmetric spaces. Such multi-component generalisations are much more complicated compared to the continuous case and, to the best of our knowledge, they were not studied up till now.

\end{itemize}

\section*{Acknowledgements} The authors have the pleasure to thank Prof. Mark Ablowitz and Prof. Vladimir Gerdjikov for numerous useful discussions.

\end{document}